# A Blockchain-based Framework for Detecting Malicious Mobile Applications in App Stores

Sajad Homayoun
Department of Computer Engineering and Information Technology
Shiraz University of Technology
Shiraz, Iran
s.homayoun@sutech.ac.ir

Ali Dehghantanha
School of Computer Science
University of Guelph
Ontario, Canada
adehghan@uoguelph.ca

Reza M. Parizi
Department of Software Engineering and Game Development
Kennesaw State University
Marietta, GA 30060, USA
rparizi1@kennesaw.edu

Kim-Kwang Raymond Choo
Department of Information Systems and Cyber Security and Department of Electrical and Computer Engineering
The University of Texas at San Antonio
San Antonio, TX 78249, USA

*Abstract*— The dramatic growth in smartphone malware shows that malicious program developers are shifting from traditional PC systems to smartphone devices. Therefore, security researchers are also moving towards proposing novel antimalware methods to provide adequate protection. This paper proposes a Blockchain-Based Malware Detection Framework (B2MDF) for detecting malicious mobile applications in mobile applications marketplaces (app stores). The framework consists of two internal and external private blockchains forming a dual private blockchain as well as a consortium blockchain for the final decision. The internal private blockchain stores feature blocks extracted by both static and dynamic feature extractors, while the external blockchain stores detection results as blocks for current versions of applications. B2MDF also shares feature blocks with third parties, and this helps antimalware vendors to provide more accurate solutions.

## I. INTRODUCTION

The estimated five billion mobile subscribers worldwide in 2017 proves the impact of mobile on people lives. Furthermore, the unique mobile subscribers are predicted to reach 71% of the world's population in 2025 [1]. According to McAfee Mobile Threat Report [2], 2017 was the year of explosion in mobile malware. Moreover, it is anticipated that 2019 will be even riskier for smartphone users, and the increasing rate of smartphone users convinced malware developers to move towards creating mobile malware. In 2017 Kaspersky detected 5,730,916 malicious installation packages, 94,368 mobile banking Trojans, and 544,107 mobile ransomware Trojans. Google as a prominent Android App store developed Google Bouncer for scanning both new applications and those already in the market [3]. However, attackers could still bypass Google Bouncer to send malware into Google's App store (Google Play) [4]. Moreover, the increasing rate of malicious apps proves inefficiency of malware detection system used by Mobile Application Marketplace [5].

In order to deliver adequate computer systems protection, antimalware solutions should be capable of detecting a very wide range of existing malicious programs [6] as well as detecting new modifications of known malware samples or new zero-day malicious programs from a recent new malware generation [7].

Recently and after the success story of blockchain on financial applications, it is rapidly entering into various applications such as healthcare, Internet of Things (IoT), Smart Contracts and Governmental applications [8], [9] A blockchain is a shared network of databases (distributed ledger) spread across multiple entities that facilitate the process of recording transactions and tracking assets [10]. There are three types of blockchain:

- Public blockchain: anyone in the world can read and send transactions, and expect to see their valid transactions into the blockchain, and anyone can participate in the consensus process.
- Private blockchain: write permissions are kept to one organization while read permissions may be public or restricted to certain participants.
- Consortium blockchain: a blockchain where the consensus process is controlled by a pre-selected set of nodes, for example, 2/3 of all participants must sign a block to be appended to the blockchain.

This paper proposes a framework for detecting malicious applications in online mobile app stores. We integrate feature extraction into a private blockchain to extract features for feeding to detection engines which determine if an application is malicious based on available feature blocks. The consortium blockchain provide mechanism of final decision by considering results of all detection engines.

The remainder of this paper is organized as follows. Section II reviews some related research. Section III describes our proposed framework for detecting malware samples in app stores. Finally, section IV discusses about the achievements of this paper and concludes the paper.

## II. RELATED WORK

Most mobile malware detection systems are focused on local file analysis [11]. Malware analysis involves two key techniques: static analysis and dynamic analysis [12]. Static analysis examines malware without actually executing it to find malicious characteristics or suspicious codes [13], while dynamic analysis (also known as behavior analysis) executes malware in a controlled and monitored environment to observe its behavior [14]. Malware detection through detecting of anomalies in battery consumption and power usage pattern [15], operating system libraries [16] are some of dynamic malware detection approaches. A recent research by Gu et al. [17] attempted to implement Consortium Blockchain for Malware Detection and Evidence Extraction (CB-MDEE) as a decentralized malware detection system based on blockchain technology. CB-MDEE framework consists of consortium blockchain and public blockchain, and includes four layers, namely the network layer, the storage layer, the support layer, and the application layer. However, CB-MDEE may not be applicable in reality as it forces users' to install a customized application for data collection.

A mobile application may be scanned in two stages. After uploading an app by developers to the app store, the security module of app store performs the first initial scan to confirm the normal behavior of the application [18]. Apps which pass the initial security check will be available for download in app stores [19]. Third party anti-malware applications perform their own scans after installation completion. Since more of these anti-malware vendors often follow almost the same approach with the first initial scan performed by app stores they may not be as useful as they claim. Therefore, improving the accuracy of detection mechanisms used by app store for the initial first scan may decrease the rate of malware victims.

## III. BLOCKCHAIN FOR MALWARE DETECTION

Fig. 1 illustrates the proposed framework for detecting malicious mobile applications in Mobile Applications Marketplaces. The structure consists of two internal and external private blockchains forming a dual private blockchain. The Internal Private Blockchain (IPB) includes all Feature Extractor (FE) components to develop and extend the Dedicated Internal Private Blockchain (DIPB) of each available mobile application.

### A. Internal Private Blockchain (IPB)

Each application is tracking with a Dedicated Internal Private Blockchain (DIPB) that follows useful information (features) showing the app's behavioral history and static information is dependent on FE components as members of DIPBs. In other words, an FE extracts valuable information during app's lifespan, and extend the related DPB by adding new blocks. Note that we consider one DIPB for each app to avoid creating a tangled blockchain due to the tremendous number of applications in the marketplace (app store). Having dedicated Private Blockchains (PB) also brings us simplicity in processing and computation of the blockchains. As shown in Fig. 1 each FE component has full access to the DIPBs through a bi-directional connection, while other nodes of the IPB have read-only access determined by one-directional arrows in the figure.

#### 1) Static Features Extractors

These features are extracted from Android's application files. Each application in Android is in .apk format and is a type of archive file, specifically in zip format packages based on the JAR file format. The MIME type associated with APK files is application/vnd.android.package-archive. The .apk file comprises both code and resources of file just similar .jar files. Android packages contain all the necessary files for a single Android program and encapsulate valuable information that can help in understanding an application's behavior.

Fig. 2 depicts major partitions of an APK file while Table 1 describes each entry of an android APK file.

The AndroidManifest.xml file available in APK file provides the application's package name, version components and other metadata (see Table II).

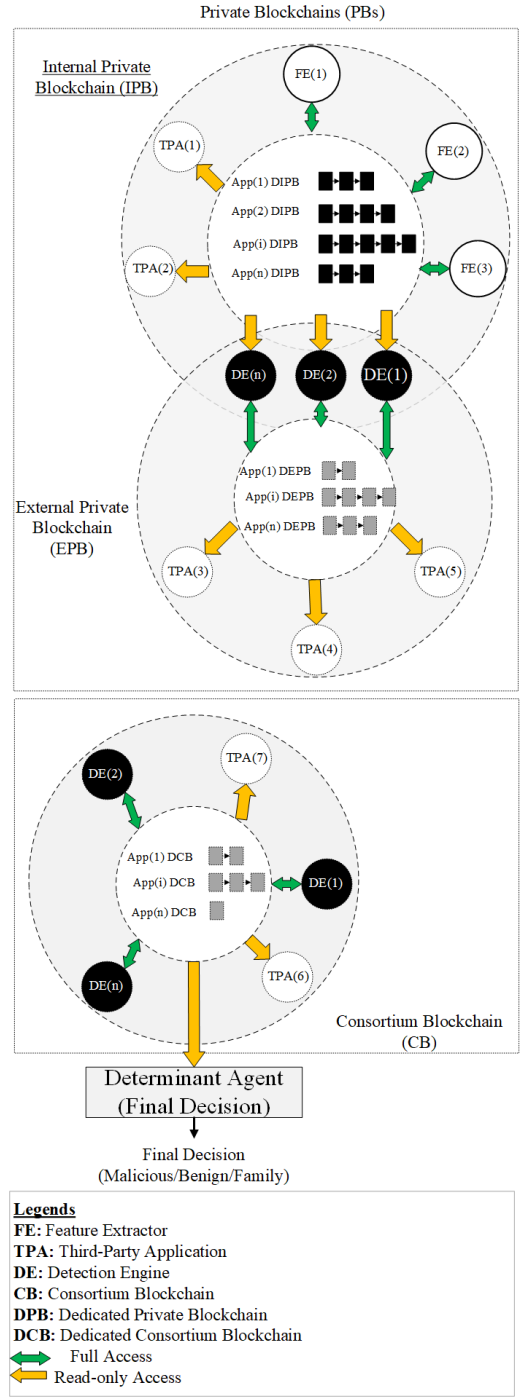

Fig. 1. The architecture of our proposed blockchain-based mobile malware detection framework (B2MDF).

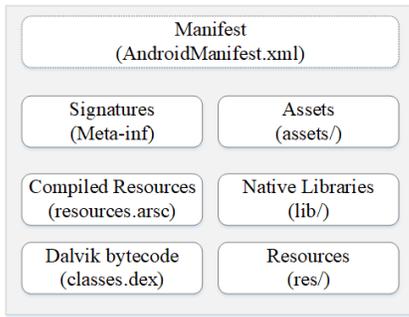

Fig. 2. An APK file structure

TABLE I. ANDROID APK FILE STRUCTURE DESCRIPTION

| Entry | Notes |
|---|---|
| AndroidManifest.xml | the manifest file in binary XML format. |
| classes.dex | application code compiled in the dex format. |
| resources.arsc | file containing precompiled application resources, in binary XML |
| res/ | folder containing resources not compiled into resources.arsc |
| assets/ | optional folder containing applications assets, which can be retrieved by AssetManager. |
| lib/ | optional folder containing compiled code - i.e. native code libraries. |
| META-INF/ | folder containing the MANIFEST.MF file, which stores meta data about the contents of the JAR. which sometimes will be store in a folder named original. The signature of the APK is also stored in this folder. |

TABLE II. SOME DETAILS OF ANDROIDMANIFEST.XML

| Attributes | Notes |
|---|---|
| Manifest tag | contains android installation mode, package name, build versions |
| Activity | Declares an activity that implements part of the application visual user interface. |
| uses-feature | Declares a single hardware or software feature that is used by the application. |
| uses-permissions | requests a permission that must be granted in order for it to operate, full list of permission API can be found in [20]. |
| Permissions | custom permission and protection level |
| Application | The declaration of the application. Will contains all the activity |
| intent-filter | Specifies the types of intents that an activity, service, or broadcast receiver can respond to. |
| provider | Declares a content provider component. A content provider is a subclass of ContentProvider that supplies structured access to data managed by the application. |
| receiver | Broadcast receivers enable applications to receive intents that are broadcast by the system or by other applications, even when other components of the application are not running. |
| service | Declare a service as one of the application components. |

Therefore, according to Table I and Table II, the following components may provide the static analysis of APK files:

- Opcode sequence FE component: Used to extract sequence of opcode from APK file to provide opcode analysis for Detection Engines (DE) of Consortium Blockchains.
- Permission FE component: Used to detect permissions requested at runtime as declared in the Manifest file.
- API calls FE component: Used to detect use of API's e.g. Telephony Manager APIs for accessing IMSI, IMEI, sending/receiving SMS, listing/installing other packages etc.
- Commands FE component: Used to detect references to system commands e.g. 'chmod', 'mount' '/system/bin/su' 'chown, etc.

*2) Dynamic Features Extractors*

Dynamic analysis based techniques attempt to detect malware applications by monitoring the runtime behavior of an app to extract useful behavioral features.

*a) System Call FE*

System calls traces often used by debuggers to control processes can be a valuable material for dynamic analysis. System calls such as file access, network connection, inter-process communication, or privilege escalation are the most common calling traces used by dynamic analyzers. "open, read, write, fork, fstat, mprotect, read, fork, write, close" is an example of a system call trace with 10 sequential activities.

*b) Memory and CPU FE*

All the features related to memory and CPU that can be accessed in Android. 53 features can be extracted for each running android application. Five CPU related features and 48 memory related features are listed in [21]. Total CPU usage, user CPU usage, and Kernel CPU usage are examples of CPU related features, while total heap size, total heap free and total heap allocated are sample features for memory.

Recall that all or a few of the mentioned Static and Dynamic FEs may contribute to DIPBs of each application as a member of the IPB. As demonstrated in Fig. 1, there are other nodes in IPB with read-only access such as third parties anti-malware applications and DE agents contributing to the DEPBs that need the information (features) provided by FECs of IPB. Accessing the information blocks facilitates third parties to contribute in scanning applications while it also provides competition between anti-malware vendors to develop more accurate detection methods.

*B. External Private Blockchain (EPB)*

Instead of DIPB in IPB, there is a Dedicated External Private Blockchain (DECB) for each application containing scanning information of different versions of applications namely malice scores assigned by each DE to each version of an application. Finally, the data stored in DECB of each application shows the history and summaries of DE's scanning results by each DE. The Detection Engines (DEs) benefits separate detection mechanisms utilizing special methods and algorithms for distinguishing malware samples from benign samples. Two basic DEs can be defined as the following:

*1) Artificial Intelligent-based DEs*

Recent developments in intelligent detection mechanisms based on machine learning and artificial intelligence provide the opportunity of detecting new unseen malicious applications [22]. There are several powerful algorithms and methods for separating malware samples from normal apps that can be used as DEs of the EPB. Each DE may perform their desired preprocessing stage on stored data in DIPBs of IPB to provide inputs for the machine learning task.

*2) Signature-based DEs*

Signature-based detection mechanisms proved that they provide robust methods for detecting malware samples based on one or more tokens or signatures. In a pretty simple form, a signature base mechanism may check hash codes of APK files to determine whether a testing sample is malicious.

## C. Consortium Blockchain (CB) and Determinant Agent

Each DE is also an active participant of the Consortium Blockchain (CB). DEs may classify an app as malicious or benign according to shreds of evidence available in EPB's DEPBs. DEs append their decision and considered pieces of evidence to DEPBs of apps. A decision can be achieving classification results (or malicious score) for an app. The CB assists Determinant Agent for the final decision showing whether an app is malicious. When a DE wants to announce an application as malicious or benign, it attempts to append a new block to the app's CB. Recall that according to previously agreed consensus policy of the CB, a block (decision information) must be accepted (signed) by a wide number of participants e.g. 2/3 of all participants must validate the decision. The CB in Fig. 1 shows that there may be other nodes with read-only access. In fact, third-party applications may want to check and use the blockchain of DE decisions for their anti-malware solutions. Finally, The Determinant Agent considers the dedicated CB of an application to determine if an application is malicious, and may also give detailed information about the type of the malware such as malware family, payload type, etc.

## IV. CONCLUSION

This paper proposed B2MDF as a malware detection framework based on a tailored-made blockchain architecture to detect malicious mobile applications in app stores before downloading by final users. Combining static analysis and dynamic analysis as an integrated solution for malware detection, in most cases, decreases the false positive rate of detection systems. The framework uses general detection engines, and it means B2MDF does not limit the implementation to any specific machine learning algorithms. B2MDF also provides useful features for third parties to develop their antimalware solutions as app stores may be more accurate in extracting features from a new uploaded sample.